\documentclass[sigconf, authorversion, noacm]{acmart}
\AtBeginDocument{%
  }

\copyrightyear{2026}
\acmYear{2026}
\setcopyright{cc}
\setcctype{by}
\acmConference[UIST '26]{The 39th Annual ACM Symposium on User Interface Software and Technology}{November 02--05, 2026}{Detroit, MI, USA}
\acmBooktitle{The 39th Annual ACM Symposium on User Interface Software and Technology (UIST '26), November 02--05, 2026, Detroit, MI, USA}
\acmDOI{10.1145/3830398.3830515}
\acmISBN{979-8-4007-2856-3/2026/11}

\usepackage{subcaption}
\usepackage[bottom]{footmisc}

\usepackage[utf8]{inputenc}

\newcommand{\flowlabel}[2]{%
  \setlength{\fboxsep}{2pt}%
  \colorbox{#1}{%
    \color{white}\textsf{\textbf{Flow: }#2}%
  }%
}

\definecolor{itoBlue}{RGB}{0,0,0}
\definecolor{itoRed}{RGB}{214,45,32}
\definecolor{itoGreen}{RGB}{0,135,68}
\definecolor{itoYellow}{RGB}{255,167,0}
\definecolor{itoViolet}{RGB}{124,58,237}
\definecolor{itoTeal}{RGB}{0,180,182}
\definecolor{itoOrange}{RGB}{255,122,25}
\newcommand{\blue}[1]{{\color{itoBlue}{#1}}}

\usepackage{listings}

\lstnewenvironment{promptbox}
  {\lstset{
    basicstyle=\ttfamily\footnotesize,
    breaklines=true,
    columns=flexible,
    backgroundcolor=\color{gray!10},
    frame=single,
    rulecolor=\color{gray!30},
    framerule=0.5pt,
    framesep=6pt,
    xleftmargin=8pt,
    xrightmargin=8pt,
    aboveskip=6pt,
    belowskip=6pt,
  }}
  {}

\usepackage{float}  
\usepackage{caption}  

\begin{document}


\title[Ito]{ITO: Real-time Browser Tab Orchestration\\ Through Intent Detection}

\author{Elvin Hu}
\affiliation{%
 \institution{Graduate School of Media Design}
 \institution{Keio University}
 \city{Yokohama}
 \country{Japan}}
 \email{elvin@keio.jp}

\author{Jasmine Sumpter}
\affiliation{%
 \institution{Graduate School of Media Design}
 \institution{Keio University}
 \city{Yokohama}
 \country{Japan}}
 \email{jasumin@keio.jp}

\author{Takatoshi Yoshida}
\affiliation{%
 \institution{Graduate School of Media Design}
 \institution{Keio University}
 \city{Yokohama}
 \country{Japan}}
 \email{yoshida@kmd.keio.ac.jp}

\author{Mauricio Sousa}
\affiliation{%
 \institution{Graduate School of Media Design}
 \institution{Keio University}
 \city{Yokohama}
 \country{Japan}}
 \email{mauricio@kmd.keio.ac.jp}

\author{Kouta Minamizawa}
\affiliation{%
 \institution{Graduate School of Media Design}
 \institution{Keio University}
 \city{Yokohama}
 \country{Japan}}
 \email{kouta@kmd.keio.ac.jp}

\renewcommand{\shortauthors}{Author et al.}

\begin{abstract}
People routinely interleave activities while browsing the web, often simultaneously and with overlapping boundaries. 
Yet organizational primitives in modern browsers treat every tab uniformly, offering no structural awareness of which items serve which purpose. 
While task-based organization approaches exist, they typically require users to manually organize or invoke reorganization features, and quickly fall out of sync as user intents evolve. 
To address this, we present Ito, a mixed-initiative approach that infers user intent in real time and organizes browsing activities into dynamic collections called Flows. 
Ito continuously reads unfolding user context, determines moment-to-moment focus, and creates, restructures, hibernates, and awakens Flows in real time while preserving user control through a mixed-initiative loop of proposals and corrections. 
In a controlled lab study (N=12) and a two-week field study (N=11), results suggest that Ito's intent-driven Flows aligned more closely with participants' mental models, facilitated task switching and navigation more effectively, and \blue{reduced management} burden without loss of perceived control.
\end{abstract}



\begin{teaserfigure}
  \includegraphics[width=\textwidth]{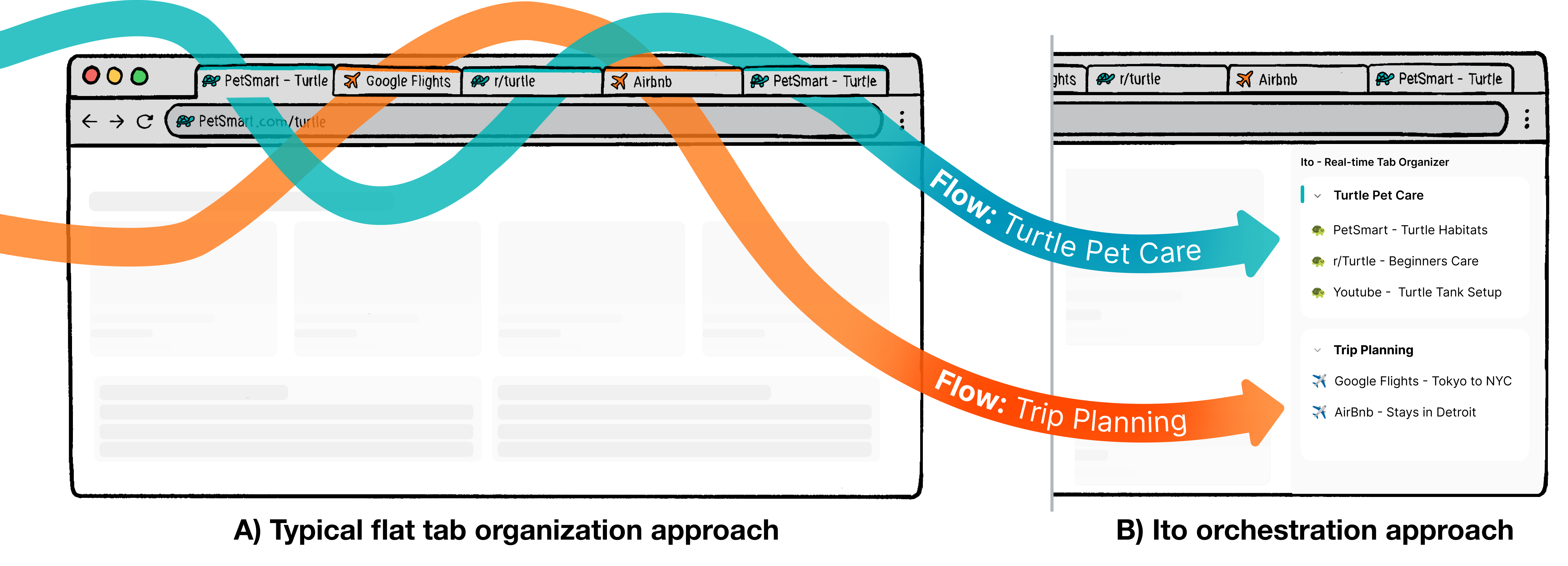}  
  \caption{
  Ito is a novel approach that organizes concurrent browsing activities into intent-driven \textit{Flows} in real time. 
  A) In a typical flat tab strip, tabs from unrelated activities accumulate in an undifferentiated row, with no structural representation of the user's concurrent goals. 
  B) Ito continuously infers the user's moment-to-moment intent and organizes tabs into \textit{Flows} in a persistent sidebar. 
  \setlength{\fboxsep}{2pt}%
  The \flowlabel{itoTeal}{~Turtle Pet Care} includes turtle-related tabs (PetSmart, r/turtle, YouTube), while the \flowlabel{itoOrange}{~Trip Planning} includes travel-related tabs (Google Flights, Airbnb). 
  These Flows reflect the user's mental model without manual organization.
  }
  \Description{
Two browser panels. Left: five mixed tabs in a flat strip, with arrows separating turtle-care tabs from travel tabs. Right: Ito's sidebar organizing those same tabs into two named groups ("Turtle Pet Care" and "Trip Planning").
  }
  \label{fig:teaser}
\end{teaserfigure}


\maketitle

\section{Introduction}

People rarely engage in single, discrete tasks when browsing. 
Instead, they navigate multiple ongoing activities that are driven by evolving underlying intents~\cite{huang2012}.
These activities unfold in parallel, frequently interrupt one another~\cite{rajamanickam2010, mark2005}, and shift as user intent changes over time~\cite{ruotsalo2014}, resulting in fluid, overlapping boundaries between pages that are difficult to represent and manage within existing browser structures~\cite{chang2021whentabscomedue,chang2021tabsdo}. 
The organizational primitives available to users (windows, tabs, file systems, app switchers) treat every item uniformly, offering no structural awareness of the purpose each serves in relation to a user's higher-level intents~\cite{bardram2006}.
As tabs accumulate rapidly across unrelated activities, people routinely end up maintaining dozens of open tabs for distinct purposes~\cite{kulkarni2019, chang2021whentabscomedue, ma2023}. 
Figure~\ref{fig:teaser}A depicts a typical browsing session in which tabs from two unrelated activities accumulate in a single undifferentiated strip, with no structural representation of the user's concurrent goals.
\blue{We refer to a user's browsing goals as \textit{intents}, which can take the form of long-term aims or moment-to-moment motivations. Unlike tasks, intents may be implicit, short-lived, interleaved with other goals, and may not cleanly map onto persistent task structures.}

Prior work has explored task- and activity-based organization as an alternative~\cite{morris2008, kulkarni2019, wang2010}. 
However, these approaches typically rely on manual grouping, requiring users to create, label, and maintain these structures themselves. 
Such effort is often perceived as not worthwhile~\cite{tashman2012lessonslearnedwindowscape, haraty2017, ma2023}, 
and even when performed quickly, it becomes outdated as user intent evolves~\cite{chang2021tabsdo}. 
At the same time, existing automated approaches remain limited in scope, often relying on surface-level signals such as domain similarity~\cite{hisatomi2025}. 
As a result, organization becomes an additional task users must manage alongside their primary goals, rather than a capability the system aids.
These limitations point to a fundamental mismatch between browsing representation and usage behavior: 
while user intent is dynamic and continuously evolving, existing browser structures remain static or require manual intervention.

In this paper, we introduce Ito, a mixed-initiative interaction approach that leverages intent-aware orchestration to organize interleaved browsing activities into dynamic collections called Flows. 
Ito continuously infers user intent in real time and restructures the browsing context accordingly (as depicted in Figure~\ref{fig:teaser}B), while preserving user agency through a loop system of proposals and user corrections. 

We evaluated Ito through two studies: 
(1) a controlled lab study comparing it against two baselines: manual tab grouping in Google Chrome\footnote{\textbf{Google Chrome:} \href{https://www.google.com/chrome}{https://www.google.com/chrome}} and the user-invoked AI-assisted organization in ChatGPT's Atlas browser\footnote{\textbf{ChatGPT Atlas:} \href{https://chatgpt.com/atlas/}{https://chatgpt.com/atlas/}}; 
and (2) a two-week field study capturing real-world browsing behavior. 
Results suggest that Ito's intent-driven Flows aligned more closely with participants' mental models than either baseline, which did not differ from one another. 
Flows facilitated task switching and direct navigation more effectively than both alternatives, suggesting that Ito can reduce the burden of tab management. 
Notably, participants reported no loss of control despite Ito autonomously restructuring their workspace. 
Together, these findings suggest that Ito can improve how people organize, navigate, and manage concurrent browsing, aligning workspace structure with evolving intent while preserving agency.

Therefore, this paper makes the following contributions:
\begin{itemize}
    \item Ito, a novel interaction approach using intent-aware orchestration for organizing concurrent user activities.
    
    \item A prototype implementing Ito through a routing pipeline that assigns webpages to Flows in real time.
    
    \item \blue{Results from a controlled lab study and a two-week field study, suggesting that intent-driven Flows are perceived more positively than both manual tab grouping and user-invoked AI grouping for organizing concurrent activities.}
\end{itemize}

\section{Related Work}

In this section we review prior work demonstrating that web browsing is often multi-threaded. 
We then summarize task-based organization systems and the mixed-initiative strategies they adopt, highlighting their benefits and limitations. 
Lastly, we review emerging intent-based approaches and identify two gaps: modeling intent continuously as browsing unfolds, and supporting the lifecycle of organized groups after their creation.

\subsection{Multi-Threaded Activity}
Modern web browsing is inherently multi-threaded. 
People frequently interleave focused work, quick micro-tasks, and leisure browsing within the same session~\cite{crichton2021, huang2010}.
Dubroy and Balakrishnan's two-week diary and click-stream study showed that people open tabs to fork exploration paths, buffer intermediate results, and multitask across concurrent threads~\cite{dubroy2010}. 
Across logs from over 50 million users, Huang and White~\cite{huang2010} demonstrated that parallel browsing is the dominant mode of web use, and subsequent work by Huang et al.~\cite {huang2012} showed that users actively rely on tabs to maintain multiple concurrent information threads.
However, as tab collections grow, managing them becomes increasingly difficult. 
Large tab sets introduce significant organizational overhead and cognitive load, leading people to adopt ad hoc coping strategies~\cite{chang2021whentabscomedue, ma2023}.
As they pursue multiple goals simultaneously, the relationships among pages become even harder to manage~\cite{hahn2018, chang2021tabsdo}. 
This mismatch between multi-threaded usage and single-threaded representation has persisted for decades, both in browsing and across computing~\cite{bannon1983, gonzalez2004, mark2005}. 
Our work addresses this mismatch by introducing an interaction approach that continuously surfaces and organizes latent threads, dynamically structuring browsing activity to align with users' evolving tasks and intents. 

\subsection{Task-Oriented Organization and Mixed-Initiative Interfaces}
To address the challenges of multi-threaded browsing, researchers have explored organizing browsing around tasks rather than individual pages. 
This line of work draws on activity-based computing, informed by activity theory, which sought to mitigate information fragmentation by structuring digital work around user activities rather than applications or documents~\cite{kuuti1995, bardram2006, bardram2019}. 
Prior approaches have adopted different strategies for determining which items belong together, which we organize below along the spectrum used by Tashman and Edwards~\cite{tashman2012lessonslearnedwindowscape} to characterize group-creation approaches.
At one end, systems require users to explicitly create and maintain groups: early desktop environments introduced virtual workspaces for this purpose~\cite{henderson1986}, 
and browser adaptations followed with tools for grouping related pages to support task switching and interruption recovery~\cite{morris2008, rajamanickam2010, kulkarni2019}, 
mirroring similar efforts in domains such as task-centric email management~\cite{bellotti2003}. 
These approaches maximize control but impose ongoing organizational effort, 
a form of "meta-activity" or "metawork" that competes with the user's primary task~\cite{tashman2012lessonslearnedwindowscape, gonzalez2004}. 
Empirical evidence suggests that many users find this overhead not worthwhile~\cite{chang2021whentabscomedue}, 
consistent with broader tendencies to accumulate digital artifacts rather than organize them~\cite{vitale2018}.

At the other end, systems such as UMEA~\cite{kaptelinin2003}, TaskTracer~\cite{dragunov2005} and Tabs.do~\cite{chang2021tabsdo} attempt to infer groupings automatically from interaction histories. 
More recent work has also explored inferring tab relationships from browsing behavior, link structure, or page content~\cite{hisatomi2025, dipinkp2019}. 
However, users are often intolerant of errors in automated organization~\cite{haraty2017}, 
and the effort to clean up incorrect groupings can outweigh the perceived benefits of automation~\cite{tashman2012lessonslearnedwindowscape, bardram2006}. 
To address this, systems often integrate mixed-initiative strategies, which are characterized by a shared control model between the user and the system, allowing human intelligence to guide automated inferences~\cite{horvitz1999}.
This approach is central to interactive machine learning, because it empowers users to directly steer model behavior, incrementally correct errors, and rapidly adapt the system to their unique workflows without needing programming expertise~\cite{amershi2014}. 
However, as studies employing this approach point out, relying on continuous user feedback introduces caveats. If a system constantly solicits corrections from users, it imposes a high interaction cost on users, which can frustrate them and cause them to ignore the system entirely if the perceived effort outweighs the benefits~\cite{kulesza2015, cakmak2012}.
Shen et al.'s study corroborates this in the context of organization, finding that prompting users for explicit corrections can incur a high interruption cost, even if the system already takes on parts of the organization~\cite{shen2009}.

Between these two ends, some systems, such as WindowScape~\cite{tashman2012lessonslearnedwindowscape}, derive groupings from implicit behavioral signals, but they fail to accommodate diverse usage styles and introduce new problems at retrieval time.

Across this spectrum, a trade-off between automation and control persists: more automation means more errors to correct, and more control means more burden to bear. Rather than seeking a better position on this spectrum of organizing based on tasks, our work organizes browsing around a more fluid signal: intent.

\subsection{Intent-Based Organization}
Whether explicit, inferred, or implicit, the approaches surveyed above share a deeper assumption that browsing activity can be meaningfully organized around tasks that are well-defined and relatively stable over time. 
In practice, users frequently shift between intents within the same session, such as comparing products, following a line of research, quickly checking a fact or watching videos for leisure. 
More recent work has begun to explore intent as a more fluid alternative. 
Systems such as IntentPrism apply AI models to infer and visualize users' information-seeking goals during browsing~\cite{wang2025}. 
However, they model intent retrospectively or at discrete points in time~\cite{mitsui2016, hisatomi2025}, 
rather than support an ongoing browsing journey and the shifting between intents.

Organizing around intent also opens a path to addressing another gap in prior work. 
Research in this space has focused almost entirely on how groups are created, not on what happens to them afterward. 
With rare exceptions outside the browsing domain, such as detecting task completion in digital assistants~\cite{white2019} or undeclared activity switches on the desktop~\cite{shen2009}, 
managing the lifecycle of organized groups remains largely unexplored. 
Tashman and Edwards' research reinforces this: participants valued peripheral awareness and retrieval of their windows far more than the grouping features themselves~\cite{tashman2012lessonslearnedwindowscape}. 
Salvucci et al.'s~\cite{salvucci2009} unified theory of multitasking, synthesizing the ACT-R cognitive architecture~\cite{anderson2007}, threaded cognition~\cite{salvucci2008}, and memory-for-goals~\cite{altmann2002}, establishes that the cognitive costs of managing concurrent activities compound over time as task context decays and resumption grows more effortful. These costs motivate lifecycle support beyond group creation: resurfacing deferred work before its context is lost, helping users maintain focus relative to an ongoing intent, and detecting when an activity may be complete.

Our work addresses both gaps. We adopt intent rather than tasks as the unit of organization, and extend support beyond group creation to encompass the full lifecycle of organized browsing activities. We use real-time LLM-based intent inference to structure browsing activity into groups as it unfolds. These groups are visible, named, and fully manipulable. The user can rename, correct, merge, or restructure them at any time. Drawing on principles of mixed-initiative interaction -- that effective human-AI collaboration requires the system to externalize its understanding so that users can inspect, modify, or reject its inferences~\cite{horvitz1999, amershi2019} -- this combination of continuous inference with persistent user control grounds our approach in the mixed-initiative tradition while organizing around a more fluid abstraction than prior task-based systems have assumed.

\begin{figure}[!b]
    \centering
    \includegraphics[width=\linewidth]{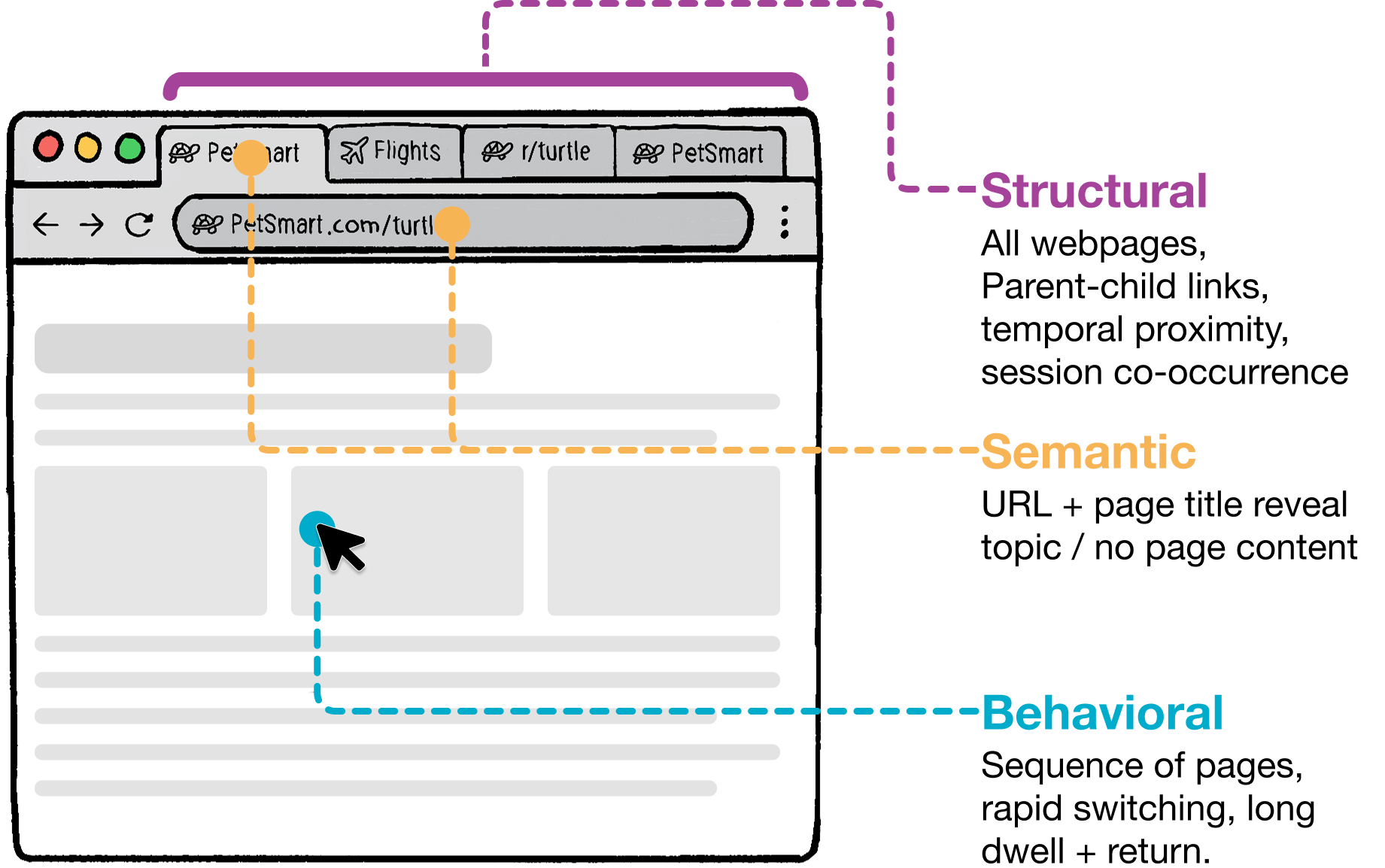}
    \caption{The three contextual signal layers Ito reads from a single browsing moment to infer user intent.}
    \Description{A browser window annotated with three dashed callouts identifying the signal layers Ito reads to infer intent: Structural, Semantic, and Behavioural.}
    \label{fig:understanding-context}
\end{figure}

\begin{figure*}[!t]
    \centering
    \includegraphics[width=\linewidth]{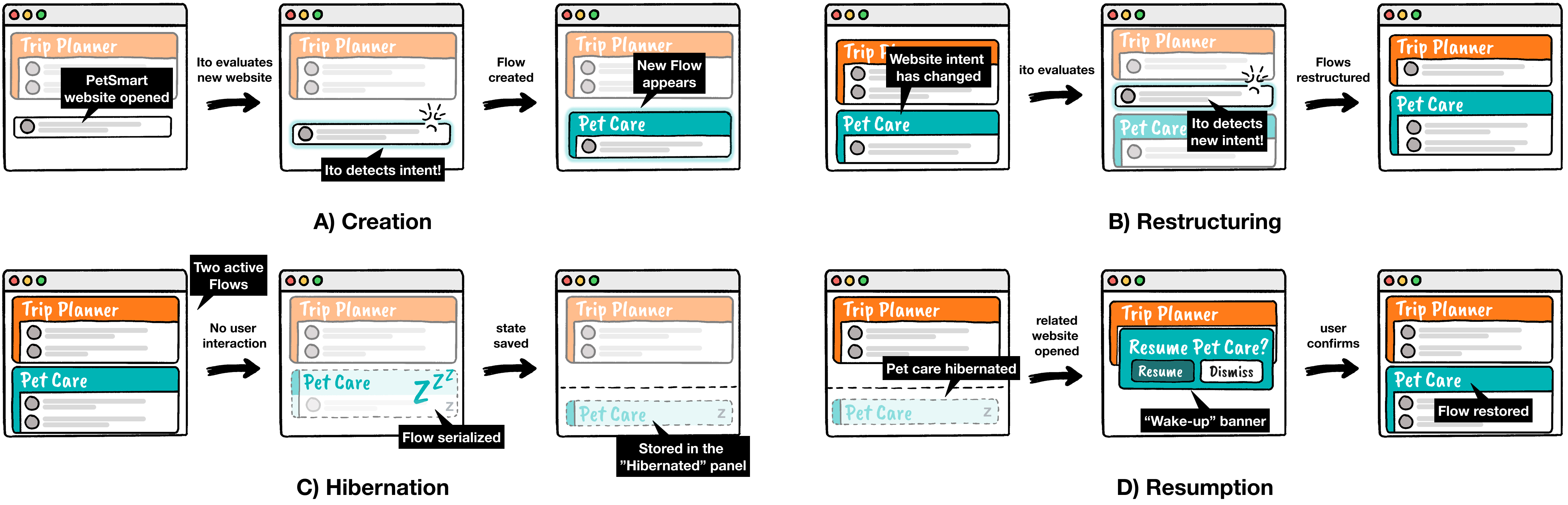}
    \caption{The Flow lifecycle in Ito. A)~\textbf{Creation:} a new intent is inferred, and a Flow is created automatically. B)~\textbf{Restructuring:} a misaligned tab is moved to a better-fitting Flow. C)~\textbf{Hibernation:} an inactive Flow is serialized and removed from the active workspace. D)~\textbf{Resumption:} related browsing triggers a wake-up prompt, restoring the Flow with its structure intact.}
    \Description{A four-part storyboard illustrating the Flow lifecycle in Ito, each part showing three browser sidebar states left to right. The four lifecycle phases are Creation, where Ito infers a new intent and creates a Flow automatically; Restructuring, where Ito moves a misaligned tab to a better-fitting Flow; Hibernation, where an inactive Flow is serialized and removed from the active workspace; and Resumption, where Ito detects related browsing and prompts the user to restore a hibernated Flow.}
    \label{fig:flow-lifecycle}
\end{figure*}

\section{Ito}
\textit{Ito}~\footnote{Ito is a Japanese homophone for both "intent" and "thread".} is a novel approach to tab orchestration based on users' intent. 
To put the problem Ito addresses in context (as depicted in Figure~\ref{fig:teaser}A), people's browsers accumulate tabs across multiple concurrent activities, such as caring for a pet turtle or planning a trip, yet the tab strip presents them as a single undifferentiated row, with no structural representation of the distinct threads of intent weaving through different subsets of tabs~\cite{chang2021whentabscomedue}.
These threads exist in the user's mental model but are not reflected by the browser: 
there is no indication of which tabs belong together, why they were opened, or how they relate to one another~\cite{chang2021tabsdo}.
As activities interleave, pause, and resume, the gap between the browser's representation and the user's mental model widens~\cite{chang2021whentabscomedue, chang2021tabsdo}, until users either abandon organization entirely or expend effort manually reconstructing it~\cite{ma2023}, only for that organization to become stale once they resume interleaved browsing.

Ito addresses this gap by continuously reading unfolding user context, inferring the user's moment-to-moment intent, and using it to dynamically organize concurrent activities into coherent groups called \textit{Flows}. 
Rather than labeling or statically grouping pages after the fact, Ito creates, restructures, hibernates, and wakes up Flows as intent evolves, keeping the user in control through a mixed-initiative loop of system proposals and user corrections.
In this section, we describe how Ito understands context, the lifecycle of a Flow, the operation of the mixed-initiative orchestration loop, and two features enabled by this continuous inference of intent.

\subsection{Understanding Context}

Ito's foundation is a continuous reading of user context: the signals from real-time interaction that enable the system to infer what the user is doing and why. \blue{Rather than asking users to declare their intent, Ito only infers it from how users browse, which is the angle this work sets out to study.}
This context is assembled from three complementary sources (Figure~\ref{fig:understanding-context}).

The first is \textbf{semantic}.
Page titles and URLs provide lightweight topical signals sufficient to distinguish ``comparing noise-cancelling headphones'' from ``planning a trip to Kyoto''.

The second is \textbf{structural}.
Pages carry implicit relationships through the paths that led to them: parent--child links, temporal proximity, and session co-occurrence all encode structure that can cluster pages into candidate activities before any semantic analysis.

The third is \textbf{behavioral}: how a user moves through their browsing environment. 
What was opened, from where, in what order, and how recently lets the system infer patterns of use. For example, rapid switching between two pages may suggest comparison, and a long idle period followed by a return may indicate resumption.

\blue{Across these three layers, Ito uses only signals available without accessing page content or form input, keeping page content on the user's device by design. We chose these signals to preserve privacy and to keep cost and latency low enough for real-time use. Semantic signals take the lead, while structural and behavioral signals help when semantic ones fall short.} 
Together, these layers let the system infer \textit{activities}: clusters of related pages that serve a common purpose.
Activities are not shown to the user; they are internal representations that form and maintain Flows.

\subsection{The Flow Lifecycle}

Flows are dynamic structures that evolve as browsing unfolds.
Each phase of their lifecycle can be initiated by Ito, the user, or both.

\textbf{Creation.}
A Flow is created when Ito infers an activity that fits no existing Flow, assigning a short yet specific label derived from the inferred intent (e.g., ``Headphone Shopping''), as depicted in Figure~\ref{fig:flow-lifecycle}A.
Users can also create Flows manually.

\textbf{Restructuring.}
The system continuously re-evaluates whether the current Flow structure remains accurate (Figure~\ref{fig:flow-lifecycle}B).
This includes renaming a Flow whose scope has shifted (e.g., from ``AirPods Max'' to ``Headphone Shopping'' to broaden the comparison), merging Flows that have converged, and splitting ones that have diverged.
Users can initiate any of these operations directly or trigger bulk re-evaluation to revise all grouping decisions at once.

\textbf{Hibernation.}
When users recognize that they want to return to an activity later but do not need it in their active workspace, they can hibernate its Flow (Figure~\ref{fig:flow-lifecycle}C).
Hibernating serializes the Flow's state (its label, member pages, and structure) and removes it from the active sidebar without discarding it.
Hibernated Flows remain accessible via a separate panel, allowing users to preserve a thread of work without the visual clutter of keeping it open.

\textbf{Resumption.}
Ito continuously monitors new pages against hibernated Flows.
If a newly visited page appears related to a hibernated Flow, the system presents a GUI dialog prompting the user to resume it (Figure~\ref{fig:flow-lifecycle}D).
The user can also resume a hibernated Flow manually at any time.
In either case, the Flow's pages are restored to the active workspace with their original label and structure intact.
This aims to address the ``out of sight, out of mind'' failure mode in which saved states become functionally invisible once removed from the interface.

\blue{These four phases are grounded in activity-centric computing~\cite{bardram2006, bardram2019}, which models how activities are created, set aside, and resumed. 
They also respond to recurring problems identified in browser user behaviors: the overhead of manual maintenance motivates continuous assistance and restructuring~\cite{ma2023}, and clutter and resumption cost motivate hibernation~\cite{chang2021whentabscomedue}.}

\subsection{Mixed-Initiative Orchestration}
\label{sec:mixed-initiative}

Based on the principles from Horvitz~\cite{horvitz1999},
Ito is a mixed-initiative approach built around two shared elements: \textit{context}, the continuous stream of behavioral, structural, and semantic signals, and \textit{interface}, the Flows that externalize the system's current understanding of the user's activities through inferring their intent.
Ito and the user can read and manipulate each of these elements, and neither party has exclusive control over any of them.
Ito reads context to update the interface; the user reads the interface to verify and correct it when necessary, while their browsing behavior continuously updates the context and implicitly affects the interface.
In turn, the interface also shapes user behavior: seeing activities organized into named Flows can nudge users toward browsing in ways that reinforce that structure, feeding back into the context.

The relationship between the user and the interface is an ongoing collaboration.
When a user moves a page between Flows, renames a Flow, or merges two, Ito treats the edit as a signal of their latent preferences. Learning directly from these natural edits allows the system to adapt its future routing to that user's preferred organizational style, without ever forcing them to explicitly describe it, a premise shared with Gao et al.~\cite{gao2024}.
As Ito becomes more reliable, users intervene less; when it makes errors, users correct them, and those corrections feed back to improve future inferences.

For this collaboration to take hold, the system should act continuously rather than on demand. 
Prior work has found that users frequently default to not engaging with on-demand organization tools, effectively making them equivalent to manual organization~\cite {seeliger2016}, 
and that requiring users to trigger reorganization imposes the very "meta-activity" it is meant to eliminate~\cite{tashman2012lessonslearnedwindowscape, kaptelinin2003}. 
Ito therefore acts continuously, maintaining an organized task space by default, so that users can direct their attention to their primary browsing activities rather than to its organization.

\subsection{Intent-Enabled Features}

A shared, live representation of user intent that both Ito and the user can read and act upon enables capabilities that extend beyond organization. 
Because Ito captures not only how pages are grouped but also why, it can reason about higher-order questions, such as whether an activity is complete or whether the user is drifting from their current focus. 
Ito demonstrates this through two features. 
First, \textit{Completion Detection} infers when a Flow's underlying activity may be finished and gently suggests that the user close or archive it. 
These inferences draw on both explicit signals (e.g., a purchase confirmation page) and implicit signals (e.g., declining engagement with a Flow's pages), and the resulting suggestions remain non-intrusive and dismissible; both user acceptance and dismissal feed back into the system to calibrate future behavior. 
Second,\textit{Focus Mode} provides intent-aware attention management by evaluating whether newly visited pages relate to the user's currently active Flow, rather than restricting access to specific domains or platforms. When a page is relevant, the system silently incorporates it into the focused Flow; when it is not, the system presents a prompt that allows the user to return to their task, continue anyway, or disable Focus Mode.
This design nudges users toward maintaining focus while preserving full agency. 

\section{Implementation}
We instantiated Ito as a Chrome browser extension.
In this section, we describe the interface design and the routing pipeline.

\begin{figure}[t]
    \centering
    \includegraphics[width=\linewidth]{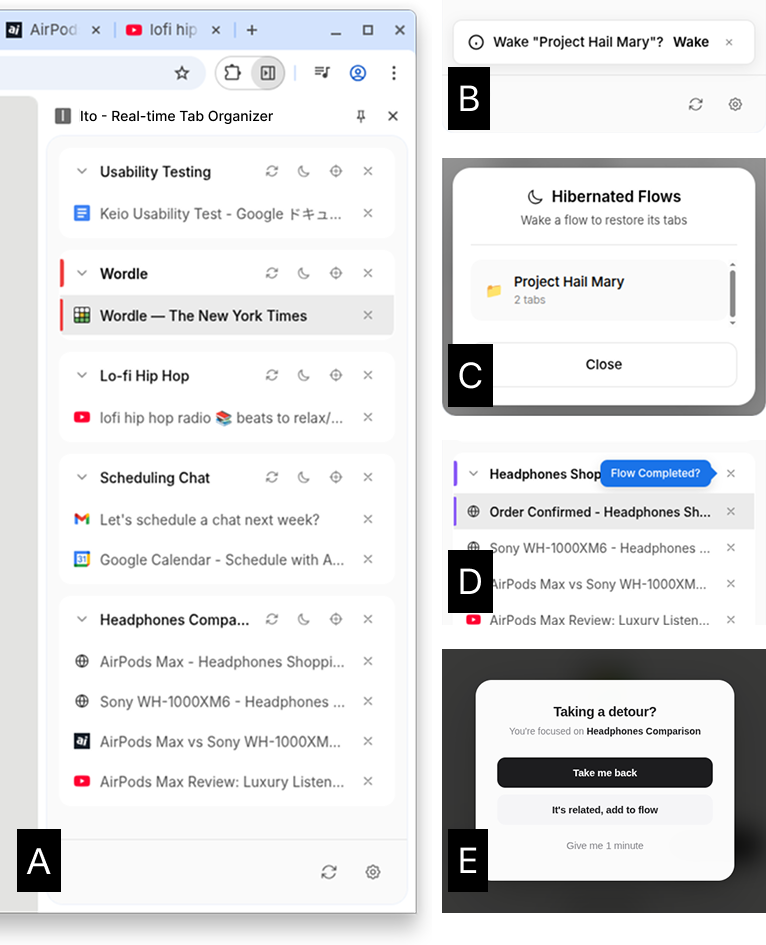}
    \caption{Ito's interface components. (A) Side panel with active Flows and member pages. (B) Automatic wake-up banner for a hibernated Flow. (C) Hibernated Flows dialog for manual restoration. (D) Completion detection prompt. (E) Focus Mode intervention modal.}
    \label{fig:UI}
\end{figure}

\subsection{UI}
Ito's interface introduces a persistent side panel integrated alongside Chrome's tab strip (Figure~\ref{fig:UI}A), which serves as the primary surface for presenting and managing Flows. 
We chose a scrollable side-panel design because it scales with concurrent Flows without sacrificing legibility. 
\blue{In an earlier version, the side panel was not scrollable as a whole and displayed all active Flows at once, with each Flow scrolling internally, so that no Flow would fall out of sight. We found that as Flow counts grew, each Flow's vertical space shrank until only one or two pages stayed visible, which made pages difficult to find.}
Placing organizational updates in the user's periphery also avoids the visual motion a horizontal layout in the focal region would introduce each time a page is routed between Flows.
Each Flow is represented as a named, collapsible group containing the pages assigned to it, labeled by activity (e.g., ``Headphone Shopping,'' ``Trip Planning'') rather than domain. 
Pages within each Flow are listed vertically by title to support quick scanning and direct navigation.
Users retain full control over Flow organization through manual operations, including dragging pages between Flows and renaming, merging, splitting, or deleting Flows. 
A dedicated ``Reorganize'' action enables users to trigger a bulk re-evaluation of all open pages, overriding the system's incremental routing. 
Manually created Flows and page assignments are treated as signals of user preference and are respected by subsequent routing decisions. 
Each Flow's header row includes controls for hibernating
\includegraphics[width=.22cm]{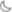}
, entering Focus Mode 
\includegraphics[width=.22cm]{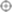}
, triggering a local re-evaluation \includegraphics[width=.22cm]{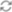}, and closing the Flow.
Hibernated Flows are removed from the active side panel but remain accessible through a dedicated dialog.

Beyond these persistent elements, the interface uses lightweight transient surfaces for context-sensitive feedback: blue bubbles for Completion Detection suggestions, banners for wake-up confirmations, and a full-page overlay in Focus Mode that flags unrelated pages while offering options to return, correct Ito's inference, or continue temporarily.

\begin{figure*}[!t]
\centering
    \includegraphics[width=.9\linewidth]{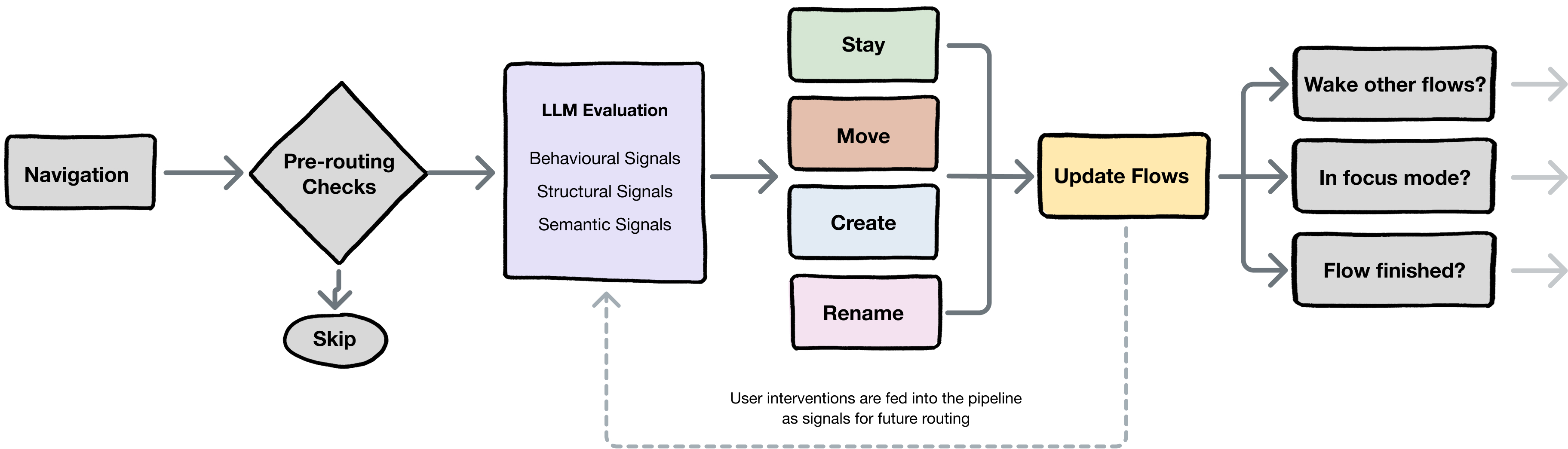}
    \caption{Routing pipeline, consisting of generally applicable steps and conditional checks.}
    \label{fig:implementation-pipeline}
\end{figure*}

\subsection{Routing Pipeline}
The routing pipeline assigns webpages to Flows in real time on every navigation event (Figure~\ref{fig:implementation-pipeline}). 
\blue{Each event first passes through deterministic checks that resolve common cases, such as redirects and login sequences, without an LLM call. We also implemented checks to prevent race conditions, such as a tab being closed while a routing call is in flight and the returned result being applied to a different tab.} Events that require a decision are then sent to an LLM.
We chose an LLM because jointly interpreting semantic, behavioral, and structural signals exceeds the capability of surface-level approaches, particularly when the same platform serves distinct intents~\cite{hisatomi2025, chang2021tabsdo}, and because user corrections can be fed directly into each routing call, personalizing decisions without retraining.
Each call includes the current page, recently visited pages, open Flows and their members, and recent user corrections.
The LLM returns one of the four decisions: \textsc{Stay}, \textsc{Move}, \textsc{Create}, \blue{or \textsc{Rename}}. 
We used GPT-4o mini in the prototype. 
\blue{We selected it after reviewing models positioned for fast, low-cost inference, narrowing to GPT-4o mini and GPT-5 nano. 
Despite being marketed as the lower-cost option, GPT-5 nano was both slower and roughly four times more expensive in our testing (50 equivalent calls) due to reasoning-token overhead. 
GPT-4o mini balanced speed and cost while producing usable routing decisions. 
Across roughly 20,000 calls during development, routing latency averaged about 2.12s at roughly \$0.00023 per call.}
We include these details to illustrate our process rather than to argue that GPT-4o mini is the ideal model for continuous intent inference; 
as models improve, it can be substituted. 
System and single-tab-routing prompts are in Appendix~\ref{app:prompts}.
\section{User Evaluation}
We conducted a remote, moderated within-subjects study to evaluate whether Ito's continuous, intent-driven organization improves mental model alignment, reduces \blue{management burden}, and supports task switching and resumption, without diminishing perceived control. 
In this evaluation, we compared Ito to Chrome's manual tab grouping and ChatGPT Atlas's on-demand AI grouping.

\subsection{Procedure}
Each session lasted approximately 60-75 minutes and was conducted via Google Meet with screen sharing. 
At the start, participants completed consent and demographic forms regarding their browsing habits and were reminded to close any sensitive tabs before screen sharing. 
The evaluation moderator then provided a brief overview of the study, explaining that participants would complete realistic browsing tasks with specified goals under multiple tab-management conditions, followed by questionnaires and a semi-structured interview.
For each condition, participants completed a set of web-browsing tasks.
The order of conditions and task sets was counterbalanced using a Latin square. 

Before using each tool, participants received a tutorial introducing its core interactions. 
The tutorials were designed to mirror one another in structure, covering the overall system, the primary ''hero'' feature (as described by the browser vendor, where applicable; e.g., the ChatGPT panel in ChatGPT Atlas), automated tab-management features (if any), and manual tab-management features.

After each condition, participants completed the same questionnaire assessing mental model alignment, task clarity, support for task switching and resumption, management burden, and perceived control. 
At the end of the session, participants participated in a semi-structured interview.

\begin{figure}[!b]
    \centering
    \includegraphics[width=\linewidth]{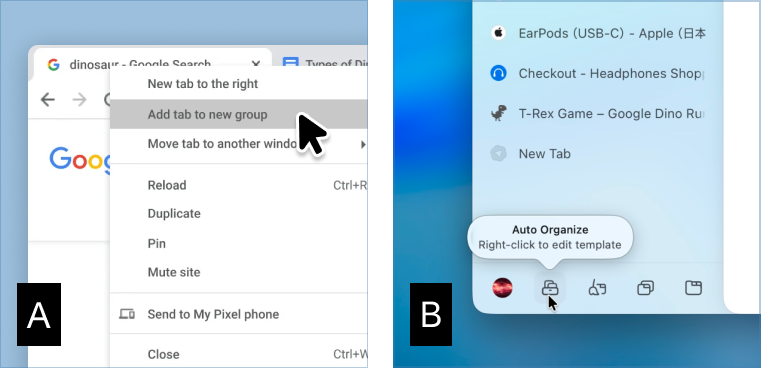}
    \caption{Baseline conditions: A) Chrome with manual tab groups; B) ChatGPT Atlas with LLM-based tab group suggestion \blue{and vertical tabs layout.}}
    \label{fig:baselines}
\end{figure}

\subsection{Tasks}
Participants performed comparable tasks such as comparing information across multiple sites, planning or coordinating across several pages, briefly shifting into leisure-oriented content, and then returning to a goal-oriented activity. 
Subtasks were categorized into 1 of 4 general activities: \textit{Shopping, Work, Ambient Entertainment, and Active Entertainment}.
These scenarios were designed to elicit the behaviors central to the following aspects: 1) whether the interface reflected how participants mentally grouped their activities, 2) whether it supported switching between subtasks, and 3) whether participants could easily resume work after interruptions.
This allowed us to compare how each tab management approach supported similar browsing demands without repeatedly exposing participants to identical content. 

To protect participant privacy, we simulated pages that would otherwise require personal information, including email, calendar, and shopping checkout; all other pages were live websites. 
Participants could complete the activities in any order. 
The complete task list is provided in the Appendix~\ref{app:tasks}.

\subsection{Setup and Prototype}
To our knowledge, Ito is the first browser tool to introduce real-time, intent-based tab orchestration, and no  single existing system can serve as a direct baseline. 
We therefore compared it against two baseline conditions: 
Chrome, which supports manual tab grouping with manual naming, serving as our baseline for fully manual organization (Figure~\ref{fig:baselines}A); and ChatGPT Atlas, which supports manual invocation with automatic naming, isolating the effect of real-time, automatic routing from on-demand AI grouping (Figure~\ref{fig:baselines}B). 

\subsection{Apparatus} %
Sessions were conducted remotely via Google Meet with screen sharing on participants' own devices, \blue{which ran macOS 14 or above as required by ChatGPT Atlas. }
Three participants \blue{whose personal Mac computers did not meet this version requirement completed the study in person on our test device while still sharing their screen via Google Meet}.
All participants were confirmed to be comfortable in their environment before the session began.

\subsection{Participants}
We recruited 12 participants (3 female, 9 male, L1-L12) aged 24-39 years ($M=29.50$, $SD=4.83$) from the university and professional communities who regularly engage in web browsing. 
Demographic questions such as age and gender were optional. 
Participants represented a range of browsing styles and tab management habits. 
We iteratively refined the prototype and study materials through 8 pilot sessions prior to the main study.

\subsection{Expert Interview}
We conducted an expert evaluation with a senior product leader experienced in managing a large-scale browser extension ecosystem and shipping consumer-facing AI features (>1B users). 
Findings from this evaluation informed the design of our lab and field studies.

\subsection{Results}
For each condition, we collected browsing data via interaction logging and subjective evaluations via the post-task questionnaire and semi-structured interview. 
Given the within-subjects design and ordinal survey data, we used the Friedman test to assess differences across conditions, followed by Wilcoxon signed-rank post hoc tests with a Holm-Bonferroni correction for pairwise comparisons.

\subsubsection{Task Performance}
All participants completed the assigned tasks within the 10-minute limit. 
\blue{To check for unintended changes in browsing behavior}, we logged four categories of navigation actions per condition (tab opens, tab closes, tab switches, and URL navigations) and recorded task completion time. Across these metrics we found no significant differences.

\subsubsection{User Preferences}
In the post-task survey, we evaluated each participant's perception of using the respective browsing tool to finish the task list. 
Friedman tests indicated significant differences across conditions for 10 of the 13 items. 
Because Chrome and Atlas did not differ significantly on any item, we report only the Ito-Chrome and Ito-Atlas comparisons below. 
We present the full results in Table~\ref{tab:field-survey}.

\paragraph{\textbf{\textit{Intent \& Task Representation (Q1--Q4).}}}
We evaluated whether users perceived each tool's layout as reflective of their mental model of ongoing activities. Friedman tests indicated significant differences across conditions for all four items: Q1 ($\chi^2(2) = 9.561$, $p = .008$), Q2 ($\chi^2(2) = 12.409$, $p = .002$), Q3 ($\chi^2(2) = 12.605$, $p = .002$), and Q4 ($\chi^2(2) = 12.842$, $p = .002$). Post-hoc Wilcoxon signed-rank tests with Holm--Bonferroni correction showed that for Q2--Q4, Ito was rated significantly higher than both Chrome (Q2: $Z = -2.917$, $p = .004$;
Q3: $Z = -2.82$, $p = .005$; Q4: $Z = -2.56$, $p = .010$) and Atlas (Q2: $Z = -2.687$, $p = .007$; Q3: $Z = -2.812$, $p = .005$; Q4: $Z = -2.825$, $p = .005$). For Q1, only the Ito--Atlas pair was significant ($Z = -2.53$, $p = .011$). 
Across Q1--Q3, Ito's IQRs (0.5, 1, and 1.25, respectively) were narrower than those of Chrome (3.25, 2.5, 3) and Atlas (2, 3, 2.25). 
For Q4, variability was comparable across conditions (Ito: 2, Chrome: 2.25, Atlas: 1.5).

\begin{table}[!t]
    \centering
    \includegraphics[width=\linewidth]{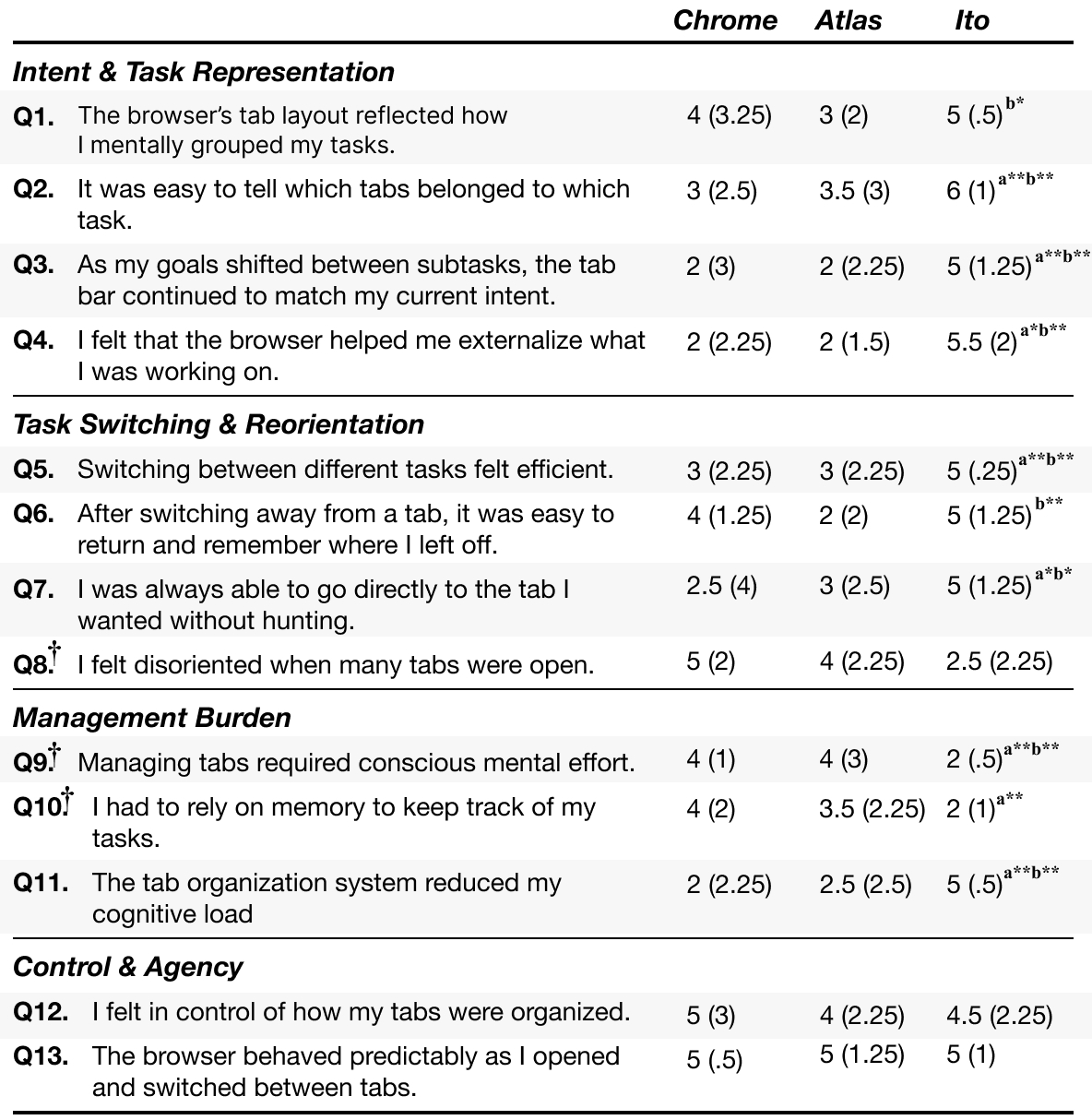}
    \caption{Post-task survey results by condition (\textit{Mdn}, IQR; $N=12$). 
$\dagger$\,=\,reverse-coded (lower is more favorable). 
Superscripts on Ito indicate significant Wilcoxon signed-rank post-hoc 
differences (Holm--Bonferroni corrected): 
$^{\rm a}$\,=\,vs.\ Chrome, $^{\rm b}$\,=\,vs.\ Atlas; 
$^{*}$\,$p<.05$, $^{**}$\,$p<.01$. 
No Chrome--Atlas pair reached significance on any item.}
    \label{tab:field-survey}
\end{table}

\paragraph{\textbf{\textit{Task Switching \& Reorientation (Q5--Q8).}}}
We assessed whether Ito's flow-based organization improved the
efficiency of switching between tasks and the ability to reorient after interruptions. Friedman tests indicated significant differences for Q5 ($\chi^2(2) = 13.227$, $p = .001$), Q6 ($\chi^2(2) = 7.302$, $p = .026$), and Q7 ($\chi^2(2) = 6.293$, $p = .043$). Q8 (disorientation, reverse-coded) did not reach significance ($\chi^2(2) = 4.732$, $p = .094$). For Q5 (switching efficiency), Ito was rated significantly higher than both Chrome ($Z = -2.949$, $p = .003$) and Atlas ($Z = -2.666$, $p = .008$), with an IQR of 0.25 indicating low variability across participants. For Q7 (direct navigation), Ito similarly received a higher rating than both Chrome ($Z = -2.355$, $p = .019$) and Atlas ($Z = -2.431$, $p = .015$). For Q6 (context recovery), Ito (Mdn = 5) differed significantly from Atlas ($Z = -2.774$, $p = .006$) but not from Chrome, which received a median of 4 compared to Atlas's median of 2. 

\paragraph{\textbf{\textit{\blue{Management Burden (Q9--Q11).}}}}
We measured whether Ito reduced the burden of managing tabs. Friedman tests indicated significant differences for all three items: Q9 ($\chi^2(2) = 15.261$, $p < .001$), Q10 ($\chi^2(2) = 7.784$, $p = .020$), and Q11 ($\chi^2(2) = 12.864$, $p = .002$). For Q9 (conscious mental effort, reverse-coded), Ito was rated as requiring significantly less effort than both Chrome ($Z = -3.126$, $p = .002$) and Atlas ($Z = -2.709$, $p = .007$); Ito's IQR of 0.5 on the 6-point scale indicates low variability, compared to Chrome (IQR = 1) and Atlas (IQR = 3). For Q10 (reliance on memory, reverse-coded), Ito scored significantly lower than Chrome ($Z = -2.616$, $p = .009$) but did not differ significantly from Atlas, though descriptive medians trended in the same direction (Ito: 2, Atlas: 3.5). For Q11 (perceived load reduction), Ito was rated significantly higher than both Chrome ($Z = -2.96$, $p = .003$) and Atlas ($Z = -2.740$, $p = .006$), again with an IQR of 0.5 compared to Chrome (2.25) and Atlas (2.5).

\paragraph{\textbf{\textit{Control \& Agency (Q12--Q13).}}}
Finally, we examined whether Ito's automatic tab routing affected perceived user agency. 
Friedman tests found no significant differences for either Q12 ($\chi^2(2) = 4.905$, $p = .086$) or Q13 ($\chi^2(2) = 2.737$, $p = .255$). Medians ranged from 4 to 5 on both items. 

\section{Field Study}

While the lab study established that Ito's intent-driven Flows were rated significantly higher than both baselines on mental model alignment, task switching, and \blue{management burden} in controlled conditions, we recognize that controlled tasks cannot fully reproduce the emergent intent structures of everyday browsing.
To understand how Ito functioned over time and to test contextual, longitudinal features such as Focus Mode and Flow Hibernation/Wake-Up, we conducted a two-week diary study with participants using Ito in their natural browsing environments.

\subsection{Procedure}

\blue{Before beginning, participants consented to the use of an external model API (GPT-4o mini) and were informed that Ito sends page titles, URLs, and Flow information, but not page content or form text. Ito also recorded basic usage information, such as which build a participant was running and whether they engaged with the features they were asked to evaluate, associated only with a participant identifier. They were reminded of this throughout the study.}
The field study used a structured diary format organized into three blocks, each introducing one feature (Focus Mode, Hibernation, Wake-Up) through a guided task followed by open-ended use during everyday browsing and concluded with a reflection survey, in which they rated their comfort with Ito and how their browsing behavior had changed compared to the previous block, using the same two questions across all three time points to track longitudinal change.
Throughout the study, participants used the same Chrome extension prototype on their personal devices in their natural working environments, with minor UI fixes applied between blocks. 
Each participant concluded with a semi-structured exit interview.

\subsection{Tasks}
Each block introduced one feature through a guided task, followed by an open-ended period in which participants were asked to engage with that feature during their everyday browsing. 

Block 1 (Day 1--5) introduced Focus Mode.
Participants first watched a short tutorial, then completed a scripted task: 
they opened two headphone-related pages, verified that Ito had grouped them into a Flow, activated Focus Mode, and searched for both related and off-topic content on Google and sites commonly associated with distraction such as YouTube and Reddit, and observed how Focus Mode responded.
After completing the task, participants were instructed to keep the Flow active and to use Focus Mode at least once during a real browsing session over the following days.

Block 2 (Day 6--10) introduced Flow Hibernation.
Participants hibernated the Flow from Block 1, manually resumed it from the hibernated Flows list, and then hibernated it again for use in Block 3.
After completing the task, participants were asked to hibernate at least one Flow of their own during everyday browsing.

Block 3 (Day 11--15) introduced the automatic Wake-Up feature. 
Participants were asked to open several pages related to the Flow they had hibernated in Block 2 and to observe whether Ito surfaced a wake-up banner prompting them to resume it.
After completing the task, participants were asked to attempt to wake a hibernated Flow at least once during their browsing, either via a banner or the hibernation panel.

\subsection{Participants}
Eleven participants (5 female, 6 male), aged 22-39 years ($M=28.27$, $SD=4.78$), of varying browsing habits completed all three diary blocks and the exit interview.
Participants were recruited from the same pool as the lab study; all had completed the lab study prior to the field study.
Participants were asked whether they would like to participate in both the lab and field studies before the lab study began. 
To reduce bias, participants who initially agreed to participate only in the lab study but later expressed interest in joining the field study were given access to the prototype and were not counted as field study participants.

\subsection{Results}

\paragraph{\textit{\textbf{Focus Mode}}}
Participants responded positively to Focus Mode as a mechanism for intent enforcement.
The feature was rated highly for helping users stay aligned with their goals (\textit{Mdn}~=~5, \textit{IQR}~=~1), and its interventions were perceived as justified by the user's own commitment to a declared intent (\textit{Mdn}~=~5, \textit{IQR}~=~1).
Participants felt in control despite the system's interruptions (\textit{Mdn}~=~4, \textit{IQR}~=~1.5) and expressed willingness to trust Focus Mode to guard high-stakes intents such as deadline-driven work (\textit{Mdn}~=~4, \textit{IQR}~=~1).
The system was not perceived as judgmental or moralizing (\textit{Mdn}~=~1, \textit{IQR}~=~1).
When asked to characterize the experience in a multi-select item, 9 of 11 participants described Focus Mode as ``the system assisting me'' and 7 of 11 as ``a reminder,'' while 4 of 11 selected ``a restriction.''

\paragraph{\textbf{\textit{Hibernation and Wake-Up}}}
All participants successfully hibernated and reopened a Flow.
They agreed that hibernation made it easier to pause one activity and switch to another (\textit{Mdn}~=~4, \textit{IQR}~=~1.5) and expressed trust in it to store important work they intended to return to later (\textit{Mdn}~=~5, \textit{IQR}~=~1.5).
The wake-up banner, which proactively surfaced a hibernated Flow when the system detected related intent, was rated highly for making it easier to resume previously unfinished activities (\textit{Mdn}~=~5, \textit{IQR}~=~1) and to recall activities that had been forgotten about (\textit{Mdn}~=~4.5, \textit{IQR}~=~1).

\begin{table}
    \centering
    \includegraphics[width=\linewidth]{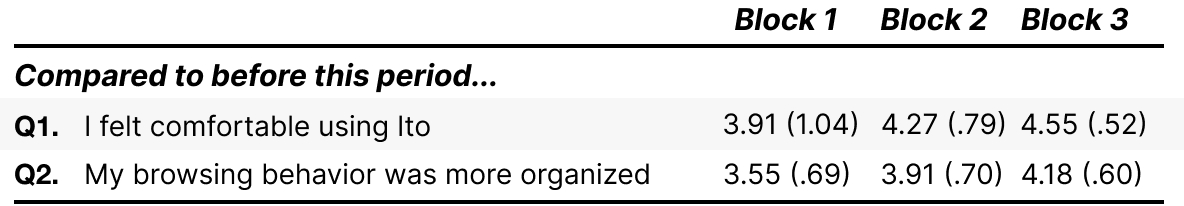}
    \caption{Reflection survey results by block (\textit{M}, SD;\textit{ N = 11}). 
}
    \label{tab:survey}
\end{table}

\paragraph{\textbf{\textit{Reflection survey (longitudinal trends).}}}
Repeated-measures ANOVAs revealed significant effects of block for both comfort ($F(2, 20) = 4.40$, $p = .026$, $\eta^2_p = .306$) and browsing behavior ($F(2, 20) = 3.89$, $p = .037$, $\eta^2_p = .280$); sphericity was not violated in either case.
Both effects were large by conventional benchmarks ($\eta^2_p > .14$).
Although pairwise Wilcoxon signed-rank tests with Holm-Bonferroni correction did not reach significance for any individual pair (all $p_{\mathrm{adj}} > .18$), the descriptive pattern suggests a monotonic increase across blocks: comfort rose from $M = 3.91$ ($SD = 1.04$) to $4.27$ ($0.79$) to $4.55$ ($0.52$), and browsing behavior from $M = 3.55$ ($SD = 0.69$) to $3.91$ ($0.70$) to $4.18$ ($0.60$).
The narrowing standard deviations further indicate that participants converged toward higher ratings over time.
The small sample size ($N = 11$) likely limited the statistical power required to survive the conservative pairwise multiple-comparison correction.
\section{Discussion}

In the lab study, Chrome and Atlas did not differ significantly on any measure.
Ito was rated more favorably than both Chrome and ChatGPT Atlas on mental model alignment, task switching, and management burden, while perceived control and predictability did not differ across conditions.
Two factors appear to contribute to these outcomes: continuous routing, and additive operations paired with low-cost corrections.
\blue{Drawing on the lab study, field study and an expert interview, we first examine these two factors.
We then position Ito among recent intent-aware interfaces, particularly in how it infers intent from implicit browsing behavior rather than explicit declaration.
Finally, we consider how Ito points to opportunities for extending mixed-initiative design principles to systems that intervene frequently and support intents that may temporarily disappear and return over time.}

\paragraph{\textbf{\textit{Continuous routing alleviates the burden that on-demand automation still imposes.}}}
\blue{Both Atlas and Ito offer AI grouping, but Atlas provides it on demand while Ito routes continuously.}
\blue{Participants rated Ito more favorably than Atlas on 9 of the 13 survey questions.
The remaining questions showed no significant differences.}
This suggests that AI grouping alone does not explain participants' preference for Ito.
We hypothesize that the difference lies in the continuity of the organization.

Atlas's on-demand grouping requires the same meta-activity as manual organization.
\blue{Users must notice clutter, decide reorganization is worth the interruption, invoke the feature, and reorient to the rearranged workspace.}

\blue{In the lab study, participants rarely invoked Atlas's on-demand grouping.
They instead defaulted to the same tab management strategy they would employ with the Chrome baseline, which did not provide any automatic tab-grouping features.
They either meticulously organized tabs by hand or simply allowed the tabs to accumulate without intervention.}
L05 explained: ``Why would I waste another five seconds to auto-group when the tab is there? I can see it.''
Even when participants used on-demand grouping, they noted that the organized state could deteriorate as soon as they navigated to a new page.
They also needed to relearn the arrangement each time Atlas reorganized their tabs.
\blue{L02 described: ``I just had to keep pressing and then every time I press it [Atlas] reorganizes. So then I have to mentally think about where each thing went.''}
\blue{The expert evaluator observed a similar pattern in other commercially available tools, noting that Chrome's tab groups, Toby, and Arc all ``require curation that users eventually abandon.''}
By contrast, Ito routes each page as it appears, allowing the workspace to remain organized without requiring users to manually intervene each time.
\blue{Because this organization happens incrementally, users have the opportunity to develop a spatial understanding of where tabs are placed, which may be more difficult to form when many tabs are reorganized at once.}
\blue{L02 described Ito's incremental nature: ``As soon as you open a new tab, it automatically sorts it to the right category.
So that helps mentally to also know, okay, this new tab just opened up and was organized here.
So that's where I can go to look for it.''}
\blue{A potential downside of Ito's routing without first requesting permission is reduced perceived control and predictability.
However, Ito received ratings similar to the baseline conditions on these two aspects (Q12 \& Q13).
We next examine the factors that may have allowed Ito to retain participants' sense of agency and control while continuously routing tabs on their behalf.}

\paragraph{
\textbf{\textit{Additive operations with multi-scale correction retain perceived control.}}}
\blue{Destructive operations remove content from the workspace, such as by automatically closing inactive tabs, while additive operations only restructure it.}
\blue{Operations in Ito are additive because it only routes tabs between Flows and never discards them.}
\blue{The expert evaluator drew an explicit distinction between additive and destructive automation.}
\blue{Grouping was described as low-risk because a miscategorized tab can be easily corrected.}
\blue{Conversely, destructive actions triggered a qualitatively different trust threshold: the evaluator stated they ``would need even stronger confidence'' before letting the system ``close off all of my tabs,'' since ``that's a destructive behavior.''}

\blue{Meanwhile, correction is available at every scale.}
\blue{Page- and Flow-level corrections are triggered by dragging, and bulk corrections are triggered by clicking a button.}
\blue{L04 described how a small correction aligned the system: ``Once I had steered it in the right direction and separated out a tab that should form a new flow, the system knew what was going on and behaved as expected afterwards.''}
When misroutings accumulate, bulk re-evaluation revises everything at once.
\blue{Bulk re-evaluation in Ito is not the same as the on-demand auto grouping feature in Atlas.}
\blue{Participants described using bulk re-evaluation in Ito as a one-time course correction, after which continuous routing maintained the result.}
\blue{With Atlas, by contrast, they had to invoke reorganization repeatedly because it was the only means of automatic organization.}
\blue{As L02 put it: ``As soon as I pressed the reorganize button, it did organize itself in a way that was coherent.''}

\blue{Notably, participants located their sense of control in knowing corrections were available, rather than in having performed them.}
\blue{L05 explained: ``I still felt in control given that I know what I can do with it.''}
\blue{L07 stated, ``The fact that I was able to move, shuffle things around is a way of feeling in control.''}
\blue{Participants organized less while retaining the ability to intervene, suggesting that additive operations with multi-scale correction sustained perceived control under continuous automation.}

\paragraph{\textbf{\textit{Inferring intent from implicit signals rather than prior declaration.}}}
\blue{Recent intent-aware interfaces have focused on helping users articulate, inspect, or directly manipulate representations of their intent.}
\blue{IntentPrism~\cite{wang2025} surfaces an intent tree from bookmarked content so users can refine goals during information foraging.}
\blue{Direct Intent Manipulation~\cite{ding2025} treats intents as first-class draggable objects on a spatial canvas, extending direct manipulation to the level of research ideation.}
\blue{These systems share a common design choice: the user must explicitly articulate their intent in language.}
\blue{Ito instead infers intent entirely from implicit browsing signals without ever requiring the user to articulate a goal.}
\blue{The resulting Flow is a legible, named representation that the user can inspect and correct, but does not need to create.}
\blue{Our studies did not directly compare this approach with explicit intent declaration, so we do not claim that implicit inference is preferable.}
\blue{Instead, Ito illustrates an alternative approach within the broader design space of intent-aware interfaces and instantiates it in browsing.}

\paragraph{
\textbf{\textit{The persistent Flow extends mixed-initiative assistance across the intent lifecycle.}}}
\blue{Ito implements many principles Horvitz established for mixed-initiative interfaces, including minimizing the cost of poor guesses and allowing users to refine automated results.}
\blue{However, these principles were developed around individual interventions within the current interaction context.}
\blue{Horvitz's examples, such as offering scheduling assistance when an email is opened, involve discrete and occasional assistive actions.}
\blue{Although the framework addresses the interruption caused by individual actions, it does not explicitly account for the cumulative impact when assistive decisions might occur after nearly every user action.}

\blue{Ito operates at a different temporal scale along two dimensions.}
\blue{First, each navigation requires a new routing decision, so system actions occur continuously rather than intermittently, and their effects may compound over time.}
\blue{Second, the intents around which Ito organizes browsing may persist for days or weeks: a headphone comparison may pause on a Tuesday and resume the following Saturday.}
\blue{The system must therefore maintain continuity not only across frequent routing decisions, but also across periods in which an intent becomes inactive and later returns.}
\blue{This combination of high-frequency action and long-lived intent creates failure modes that the original principles do not directly address:
repeated interruption if the system requests approval for every placement, and loss of continuity if the Flow disappears from the interface and is not resurfaced when the intent returns.}

\blue{Ito's later lifecycle stages, hibernation and resumption, were designed in response to these failure modes.}
\blue{Participants' ratings in the diary study provide early evidence that they valued these longitudinal features.}
\blue{However, the existing mixed-initiative principles offered little guidance for designing them.}
\blue{We present Ito's design as an initial exploration of how mixed-initiative systems might support high-frequency assistive actions around long-lived intents.}
\blue{We encourage future studies to examine what additional design considerations and principles may be needed for mixed-initiative systems operating at this temporal scale.}
\section{Limitations and Future Work}

Our mixed-methods evaluation combines controlled comparison across three conditions with a two-week longitudinal field test, providing both internal validity and ecological grounding. Still, several dimensions of the design space remain open.

\blue{Our survey used custom items rather than validated instruments. This follows common practice in browser research, where studies combine purpose-built measures with interviews and interaction logs~\cite{chang2021tabsdo, morris2008}. Validated instruments such as NASA-TLX~\cite{hart1988} could improve comparability across studies.}

The lab study's structured tasks introduced some contrivance, as participants pursued explicit goals within bounded sessions rather than the fragmented, goal-unstable browsing that characterizes everyday use.
The diary study mitigates this by observing Ito in participants' natural environments, but awareness of being studied may have influenced behavior. 
\blue{We also note that Completion Detection, while described as a potential feature that intent-based Flows could enable, was not formally evaluated in either study. In addition, our comparison evaluates continuous intent-driven organization in a browser side panel as a whole rather than isolating the contribution of routing itself. That said, layout does not appear to drive the effect: Atlas presents a vertical tab layout similar to Ito's, yet no Chrome-Atlas pair reached significance on any survey item. Still, a condition presenting the same Ito side panel without continuous grouping and routing could help isolate the effect of routing.} Future work with larger cohorts, longer study periods, and fewer researcher-imposed constraints would also help validate these findings.

We did not conduct a standalone technical evaluation of routing accuracy, given the diversity of potential browsing behaviors and personal preferences about grouping.
However, the expert evaluator described groupings as "sufficiently accurate for daily use", and field study participants corrected an average of 2.5\% of routing decisions, though this figure is likely an underestimate as participants may have tolerated borderline placements or missed misclassifications.
\blue{Future work could ask participants to label routing decisions against their own intended grouping. }

The current prototype routes page metadata through an external API, raising privacy constraints that limit both adoption and the richness of available signals.
On-device inference would eliminate this dependency, and crucially, unlock richer context such as page content.
Because Ito routes on every navigation, the trade-offs between reasoning quality, latency, cost, and privacy are sensitive to both model selection and whether inference runs on-device or server-side; future work could quantify these across configurations.

The current prototype ties Ito to a browser side panel; 
deeper integration, such as replacing the tab strip entirely, could address the space constraints and redundancy that led some participants to minimize or underuse the panel on smaller-screen devices.

Beyond the browser context, multi-threaded task management is a shared challenge across AI agent interfaces, code editors, and document tools~\cite{simkute2025, parnin2010}.
Flow-based orchestration could also extend across windows on the same device, mobile and spatial computing environments, and cross-device journeys where intent needs to persist across devices. 
\blue{The same approach could be explored in the context of the operating system, orchestrating applications by intent through surfaces like the macOS Dock or Windows Taskbar.}

Ito already captures recent Flow history and injects it into routing decisions.
Expanding it into a persistent intent record would improve routing accuracy, enable more proactive resumption, and produce a structured trail of what the user was doing and why.
Unlike tools such as iOS Screen Time that track activity without capturing the goals behind it, an intent trail could record which goals each resource served and when attention drifted, opening directions for goal-level reflection in digital-wellbeing research.
\section{Conclusion}

We presented Ito, a mixed-initiative approach that infers user intent in real time to organize browsing activities into dynamic collections called Flows. 
In a controlled lab study, results \blue{suggest} that Ito's Flows aligned more closely with participants' mental models and reduced the burden of task switching and management, without reducing perceived control.
A two-week field study extended these findings to naturalistic use, \blue{during which Ito's continuous, additive automation showed promise in} shifting users from active management to passive monitoring without loss of agency. 
Future work includes broader deployments, adaptation to other surfaces and form factors, on-device inference, and the extension of Flow semantics across devices and applications.

\begin{acks}
We thank our study participants for their time across both the lab and field studies, and our expert evaluator for their insights.
\end{acks}

\bibliographystyle{ACM-Reference-Format}
\bibliography{main}

\appendix
\section{Routing Pipeline Prompts}
\label{app:prompts}

\subsection{System Prompt}
\begin{promptbox}
[Flow] System prompt: You are a tab router for an intent-aware browser tab manager.

=== WHAT IS A FLOW ===

A flow groups tabs that serve an activity — what they're actively exploring or working on.

The flow's name is inferred from behavior, not declared. If the user has opened several bookshop tabs and review sites, their intent is "photo books" — even if they never said so. The tabs reveal the activity.

However, the user's intent across activities might shift as they browse, and your job is to detect this intent shift and route tabs to the correct flow.

Key principle: tabs that serve the SAME PURSUIT belong together, regardless of what site they're on or what specific action the user is performing (browsing, searching, comparing, buying). The pursuit is the constant; the actions might change.

=== ROUTING RULES ===

DEFAULT: STAY. The tab is already in a flow — it was placed there because it matched the flow's pursuit. Override with a clear reason. Do not hallucinate far-fetched associations.

STAY when:
- The tab's subject relates to the flow's pursuit. 
- The tab is on a related site to other tabs in the flow
- The tab is a sub-page, search, review, or comparison within the same pursuit
- You're unsure — when in doubt, STAY

MOVE to an existing flow when:
- Another flow's tabs clearly share the same specific pursuit as this tab
- Example: "iPhone 16 Review" is in "YouTube" but "iPhones" flow exists with iPhone tabs -> MOVE

CREATE NEW when:
- The tab is a COMPLETELY DIFFERENT pursuit AND no existing flow matches
- "Twitter" while in "Photo Books" -> different pursuit entirely -> NEW
- Do NOT create NEW simply because the tab is on a different site — different sites often serve the same pursuit; check the potential user intent first

RENAME when:
- The flow's name no longer captures the pursuit its tabs represent, but the tabs are still relevant to each other
- CRITICAL: Check current_flow.tabs — the new name must fit ALL tabs in the flow, not just the current tab. If the other tabs don't relate to the proposed name, use NEW instead (the current tab diverged from the flow)
- Example: "YouTube" flow where ALL tabs are about cameras -> RENAME:Cameras
- Example: Single-tab flow where content changed -> RENAME to match current content
- WRONG: Flow has "Jeremy Jordan performance" + "Camera review" -> RENAME:Camera Reviews (ignores other tab) -> should be NEW:Camera Reviews

=== SPECIAL CASES ===

- System URLs (chrome://, chrome-extension://): Never stay in a content flow -> NEW or move to utility flow, except for new tab page. New tab page should be defaulted to stay in the current flow.
- Helper sites (Google Search, ChatGPT): Route by the QUERY/SEARCH TERM, not the site — "project hail mary review reddit - Google Search" is about Project Hail Mary, not about Google. 
- User-named flows (user_named: true): Interpret scope broadly, prefer STAY
- Link-opened tabs: The parent flow is context — if the tab supports the parent's pursuit, keep it there

=== LEARNED PREFERENCES ===

Check learned_preferences.recent_corrections_summary — if the user previously moved similar tabs, follow that pattern.

=== NAMING ===

2 words max. Name by the SUBJECT of the pursuit

=== YOUR TASK ===

You will receive a JSON object with: the tab to route, its current flow, other existing flows, and signals.

=== SIGNALS ===

Use these as context hints, not hard rules:
- same_domain_nav: Browsing within same site — default to continuity, not separation, unless the subject has changed drastically
- is_link_opened / has_parent_chain: Opened from another tab — parent's flow is strong context
- first_evaluation: First time routing this tab (vs re-evaluation after content change)
- title_changed: Content updated — check if pursuit shifted
- is_new_tab_page: Transitional — prefer STAY
- is_system_url: chrome:// pages — never keep in a content flow -> NEW or utility flow
- manual_creation: User typed URL or Cmd+T — intentional navigation

=== DECISION PROCESS ===

1. What is this tab's SUBJECT? Extract from the title AND the URL path — both carry signal. 
   - Helper sites (Google, ChatGPT): route by query topic; no query yet -> STAY

2. Does the subject match the current flow's pursuit? If yes -> STAY.
   - Same domain, different pages -> usually same pursuit -> prefer STAY

3. Does the subject match another flow better? If yes -> use that flow's exact name.
   - Only move when the other flow's tabs clearly share the SAME pursuit. 
   - Do not fabricate abstract connections to force a move

4. No flow matches and the pursuit is genuinely different -> NEW:Name
   - In a single-tab flow, prefer RENAME — the tab IS the flow

=== RESPONSE FORMAT ===

Return JSON:
{"decision": "<value>", "reason": "<brief explanation>"}

Decision must be EXACTLY one of:
- "STAY" — tab fits current flow, return "RENAME:Name" when necessary
- Exact flow name from other_flows — move tab there
- "NEW:Name" — genuinely different intent/pursuit, no existing flow matches

Examples:
{"decision": "STAY", "reason": "Tab supports current intent"}
{"decision": "RENAME:Cameras", "reason": "Flow tabs now all about cameras"}
{"decision": "iPhone", "reason": "iPhone review belongs with other iPhone tabs"}
{"decision": "NEW:Twitter", "reason": "Social browsing, unrelated to current intent"}

\end{promptbox}

\subsection{Single-tab Routing Prompt}
\begin{promptbox}
{
  "signals": {
    "is_new_tab_page": false,
    "is_interstitial": false,
    "is_system_url": null,
    "domain_changed": false,
    "title_changed": true,
    "same_domain_nav": false,
    "has_parent_chain": false,
    "parent_info": null,
    "manual_creation": true,
    "first_evaluation": true,
    "is_link_opened": false
  },
  "environment": {
    "mode": "single_tab",
    "focus_mode": {
      "active": false,
      "focused_flow": null
    }
  },
  "current_flow": {
    "name": "[Current Flow Name]",
    "user_named": false,
    "stability": "[Status]",
    "tab_count": 0,
    "tabs": [
      {
        "title": "[Tab Title]",
        "domain": "[Domain]"
      },
      {
        "title": "[Tab Title]",
        "domain": "[Domain]"
      }
    ]
  },
  "other_flows": [
    {
      "name": "[Flow Name 1]",
      "user_named": false,
      "stability": "[Status]",
      "tab_count": 0,
      "tabs": [
        {
          "title": "[Tab Title]",
          "domain": "[Domain]"
        }
      ]
    },
    {
      "name": "[Flow Name 2]",
      "user_named": false,
      "stability": "[Status]",
      "tab_count": 0,
      "tabs": [
        {
          "title": "[Tab Title]",
          "domain": "[Domain]"
        }
      ]
    },
    {
      "name": "[Flow Name N]",
      "user_named": false,
      "stability": "[Status]",
      "tab_count": 0,
      "tabs": [
        {
          "title": "[Tab Title]",
          "domain": "[Domain]"
        },
        {
          "title": "[Tab Title]",
          "domain": "[Domain]"
        },
        {
          "title": "[Tab Title]",
          "domain": "[Domain]"
        }
      ]
    }
  ],
  "tab": {
    "title": "[Target Tab Title]",
    "domain": "[Target Domain]",
    "url": "[Target URL]"
  }
}
\end{promptbox}

\clearpage
\section{Lab Study Task List}
\label{app:tasks}

\subsection{Task Set A}
\begin{promptbox}
In separate tabs, open the product pages for Airpod Max and the Sony xm6 respectively from 
Airpods Max: [GitHub URL hidden. Blind for review]
Sony:  [GitHub URL hidden. Blind for review]

Open this email website and read a message about scheduling a meeting.

In a new tab, search ''Airpod Max vs Sony xm6'' and open any review article.

In a new tab, open youtube.com. Search ''airpod max 2nd gen review'' and open any review video

In a new tab, open youtube and search ''lofi hip hop radio''. Open the result called '' lofi hip hop radio beats to relax/study to'' by lofi girl and let it play in the background.

Open calendar and check your availability.

In a new tab, open youtube.com and search ''never give up clam'' and watch the result.

Draft a response proposing a meeting time. (A simple response is okay)

Return to the background media and change what's playing (e.g., skip, switch videos, or adjust).

In a new tab, search for ''Airpod Max Usb-C Cable'' and select the top result from Apple.com

In a new tab, Open Wordle from nytimes.com. Attempt to play for two turns  

Send the drafted meeting response. 

Decide which headphone model you're most interested in so far and complete the checkout process on the shopping pages you opened at the very beginning (website name: ''Headphone Shopping'').

\end{promptbox}

\subsection{Task Set B}
\begin{promptbox}
Open this mail website and read a message about scheduling a meeting.

In separate tabs, open the product pages for Airpod pro and the Bose QC Ultra Earbuds respectively from
AirPods Pro: [GitHub URL hidden. Blind for review]
Bose QC Ultra: [GitHub URL hidden. Blind for review]

In a new tab, open youtube and search ''bossa lofi radio''. Open the result called '' bossa lofi radio chill music for relaxing days'' by lofi girl and let it play in the background.

In a new tab, search ''Airpod pro vs Bose QC Ultra'' and open an editorial review article.

Open calendar and check your availability.

In a new tab, open youtube.com. Search ''Airpod Pro 2 review'' and open the resulting review video called ''AirPods Pro 2 Review: 1 Underrated Thing!'' by Marquee Brownlee.

In a new tab, open youtube.com and search ''David Lynch Ideas'' and watch the result posted by BAMorg.

Return to the background media (lofi radio) and change what's playing (e.g., skip, switch videos, or adjust).

In a new tab, search for ''Airpod case'' and select the top result 

In a new tab, Open Suika game. Attempt to play for a minute

Draft a response proposing a meeting time. (A simple response is okay)

Decide which headphone model you're most interested in so far and complete checkout process on the shopping pages you opened at the very beginning (website name: ''Headphone Shopping'').

Send the drafted meeting response.
\end{promptbox}

\subsection{Task Set C}
\begin{promptbox}
In separate tabs, open the product pages for Earpods and the Sony EX155 respectively from
EarPod: [GitHub URL hidden. Blind for review]
Sony EX155: [GitHub URL hidden. Blind for review]

Open this email website and read a message about scheduling a meeting.

In a new tab, search ''Earpod  vs Sony EX155'' and open any review article. 

In a new tab, open youtube.com. Search ''earpods review'' and open a review video (such as ''Apple EarPods USB-C Review | Sound, Features, and Comparison to AirPods'' by SoundGuys. )

In a new tab, open youtube and search ''synthwave radio''. Open the result called ''synthwave radio beats to chill/game to'' by lofi girl and let it play in the background.

Open calendar and check your availability.

In a new tab, open youtube.com and search ''Olimar Goodnight Pikmin'' and watch the result posted by Donk\textperiodcentered tk.

Draft a response proposing a meeting time. (A simple response is okay)

Return to the background media and change what's playing (e.g., skip, switch videos, or adjust).

In a new tab, search for ''apple wired headphones'' and select the top result from Apple.com

In a new tab, Open T-rex runner. Attempt to play for up to two tries. 

Send the drafted meeting response. 

Decide which headphone model you're most interested in so far and complete the checkout process on the shopping pages you opened at the very beginning (website name: ''Headphone Shopping'').
\end{promptbox}

\clearpage
\section{Lab Study Task: Simulated Web Pages}
\label{app:pages}

\begin{figure}[H]
    \centering
    \includegraphics[width=1\linewidth]{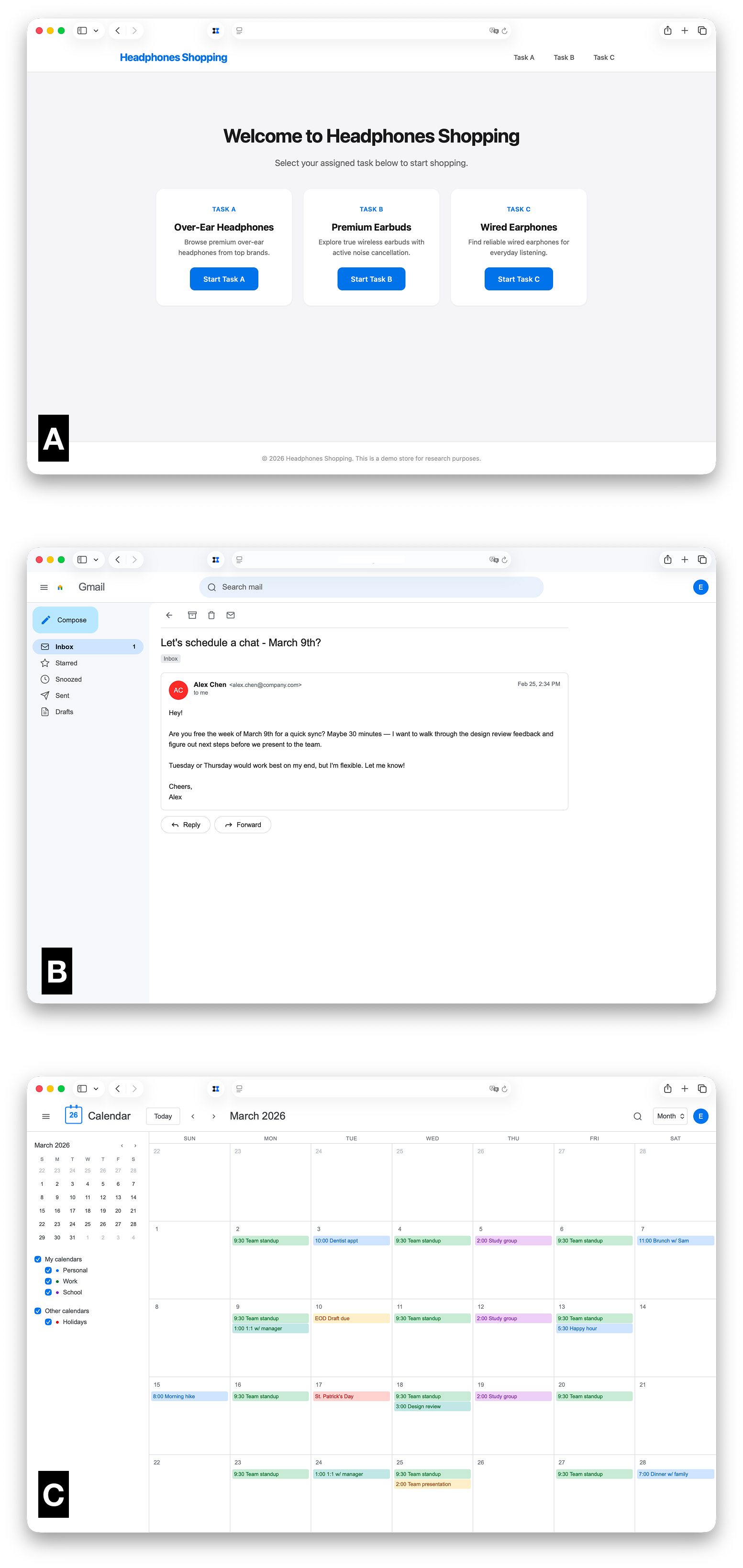}
    \caption{Simulated pages used in the Lab Study evaluation: A) Online headphones store, B) Web-mail Application, and C) Calendar Application. (All names appearing in these pages are fictitious and do not identify any author or participant).}
    \label{fig:placeholder}
\end{figure}

\end{document}